# Performance and Comparisons of STDP based and Non-STDP based Memristive Neural Networks on Hardware

*Zhiri Tang*

**Abstract**—With the development of research on memristor, memristive neural networks (MNNs) have become a hot research topic recently. Because memristor can mimic the spike timing-dependent plasticity (STDP), the research on STDP based MNNs is rapidly increasing. However, although state-of-the-art works on STDP based MNNs have many applications such as pattern recognition, STDP mechanism brings relatively complex hardware framework and low processing speed, which block MNNs' hardware realization. A non-STDP based unsupervised MNN is constructed in this paper. Through the comparison with STDP method based on two common structures including feedforward and crossbar, non-STDP based MNNs not only remain the same advantages as STDP based MNNs including high accuracy and convergence speed in pattern recognition, but also better hardware performance as few hardware resources and higher processing speed. By virtue of the combination of memristive character and simple mechanism, non-STDP based MNNs have better hardware compatibility, which may give a new viewpoint for memristive neural networks' engineering applications.

# I. INTRODUCTION

Memristor was postulated by L. O. Chua in 1971 [1] and realized by HP Labs in 2008 [2]. The value of memristor depends on the amount of electricity flowing through it [3]-[4] and the memristor has a potential on descriptions of the biological synapse's characteristics such as the spike timing-dependent plasticity (STDP) [5]-[7]. Hence, memristive neural networks (MNNs), which use memristors as biological synapses [8]-[9] in the neural computing, have become a hot research topic. Compared with Convolutional Network Networks (CNNs), MNNs give a possibility to mimic the learning rules of the biological neurons, which is very attractive for studying and exploring the neuromorphic computing [10]-[11] and the artificial intelligence [12]-[13]. Recently, various researches based on different mechanisms and structures have emerged [14]-[16]. However, it's an open question that whether STDP is the only compatible and suited method for MNNs.

Some latest works about the unsupervised memristive feedforward structure designed feedforward neural networks [17]-[18] and multilayer perceptron [19] based on STDP, which merge biological learning mechanism into memristive neural networks. Some other researches on feedforward structure [20] gave up STDP to design some new learning systems and algorithms. Some state-of-the-art works designed the unsupervised memristive crossbar structure with STDP and other biological learning rules [21]-[22] and applied them into applications including pattern recognition [23]-[24], edge detection [25]-[28], and high performance computing [29]. However, there are still many works about the non-STDP based unsupervised memristive crossbar structure [30], which can also achieve good performance in above applications.

From above state-of-the-art researches, it is generally believed that because MNNs can be based on STDP, it can better mimic biological neurons to achieve many applications and can utilize few resources and have high processing speed. However, STDP based MNNs can't meet the practical requirements of some engineering applications. Compared with STDP based MNNs, non-STDP based MNNs have better engineering compatibility and performance with the same applications and functions.

Inspired by the above, this paper builds STDP based and non-STDP based unsupervised memristive neural networks for two common memristive structures including feedforward and crossbar to carry on contrastive research, respectively. The comparison includes pattern recognition accuracy, convergence speed of training, hardware resource occupancy, and processing speed with the expansion of network scale. Through the experiments, non-STDP based MNNs show better hardware performance remaining memristive characteristics.

## II. MEMRISTIVE FEEDFORWARD STRUCTURE

In this section, a STDP based and a non-STDP based unsupervised memristive feedforward structures are built aiming to carry out experimental comparison and analysis between them.

*A. STDP based Feedforward Structure*

According to traditional HP memristor model [2], the value of memristor is:

$$M(t) = \frac{d(\phi(t))}{d(q(t))} = \frac{U(t)}{I(t)} \tag{1}$$

The relationship between the voltage at the two ends of memristor and current through memristor is:

$$U(t) = [R_{off} - (R_{off} - R_{on})\mu_v \frac{R_{on}}{D^2} \int_{-\infty}^{t} I(t)dt]I(t) \qquad (2)$$

So we have:

$$\begin{aligned} M(t) &= R_{off} - (R_{off} - R_{on})\mu_v \frac{R_{on}}{D^2} \int_{-\infty}^{t} I(t)dt \\ &= k_1 - k_2 \int_{-\infty}^{t} I(t)dt \end{aligned} \qquad (3)$$

where $k_1 = R_{off}$ and $k_2 = (R_{off} - R_{on})\mu_v \frac{R_{on}}{D^2}$

STDP is shown as Fig. 1, which indicates the relationship between the spike interval time and the percentages of synaptic weight change [5]. According to STDP, the shorter the firing time between the pre-synaptic neuron and the post-synaptic neuron, the closer the relationship between two neurons and the larger the synaptic weight between two neurons [32].

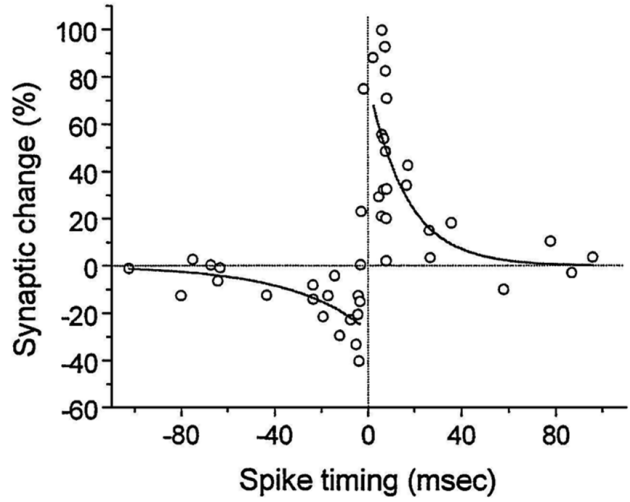

Fig. 1 Spike timing-dependent plasticity (STDP)

Due to the change of synaptic weight equaling to the change of the reciprocal value of the memristor, we have:

$$dW = dG = d(\frac{1}{M(t)}) = d(\frac{1}{k_1 - k_2 \int_{-\infty}^{t} I(t)dt}) \qquad (4)$$

which is consistent with STDP. From above we can get a STDP based memristive neural network algorithm.

Spiking response model (SRM) [32] is chosen as spiking signal [33] transmission mode. SRM builds an unsupervised learning mechanism which is similar to biological neurons with STDP. According to SRM, spikes generated by presynaptic neurons will contribute a post synaptic potential (PSP) [34] to the membrane potential of postsynaptic neuron. A general PSP definition is:

$$f_{psp} = w(e^{-\frac{t}{4}} - e^{-\frac{t}{2}}) \tag{5}$$

The schematic diagram of PSP is shown as Fig. 2, from which it can be seen that the height of spikes depends on the synaptic weight between these two neurons. When two spikes from presynaptic neuron input to postsynaptic neuron in succession, the membrane potential caused by PSP will add up. If the sum of PSP causing potential is up to a fixed threshold, the postsynaptic neuron will fire and an output spike will input to next neuron. After firing, the postsynaptic neuron will go into a refractory period, during which the neuron will not response to any spikes from presynaptic neurons. The postsynaptic neuron will restore normal response to the spikes after refractory.

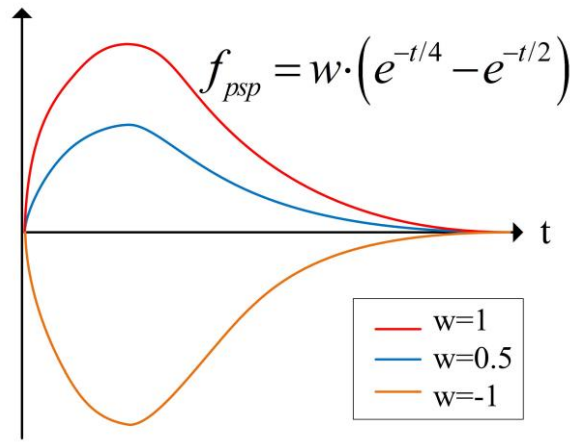

Fig. 2 Post synaptic potential

The entire SRM is shown as Fig. 3, in which Fig. 3(a) is presynaptic spikes, Fig. 3(b) is membrane potential caused by PSP, Fig. 3(c) is the sum of membrane potential, and Fig. 3(d) is output spikes of postsynaptic neuron. Since STDP mechanism is relatively complex, normally the STDP based memristive neural network use some hardware optimizations [35], which includes simplification, linearization, and pipeline design, to achieve better performance on hardware platform.

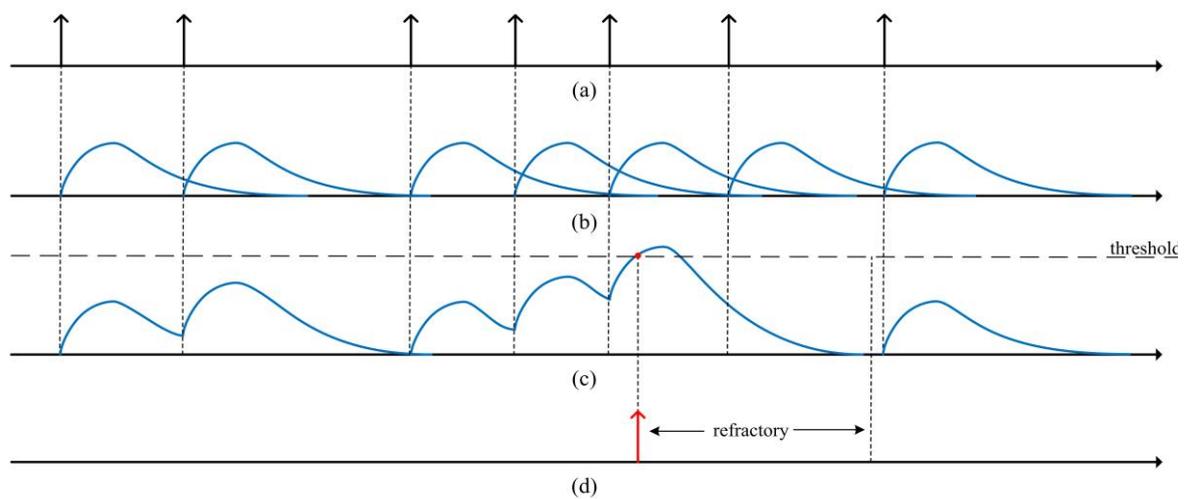

Fig. 3 Spiking response model

The entire structure of STDP based memristive feedforward neural network is shown as Fig. 4. First, original binary images are encoded into black-pixel spikes and white-pixel spikes. The black spikes have higher firing frequency than white spikes over the same period so the output spikes corresponding to black spikes will fire earlier. If the input image has 9*9 pixels, the presynaptic neuron layer will have 81 neurons. With the calculation module including PSP and SRM, output spikes of postsynaptic neuron layer can be got. The input spikes and output spikes are added to two ends of memristors, respectively. Then the feedforward networks will process training and weight updating according to STDP unsupervised learning mechanism of memristors. Further, a fixed threshold for memristors is set to make sure that training process will end after the values of memristors exceed it.

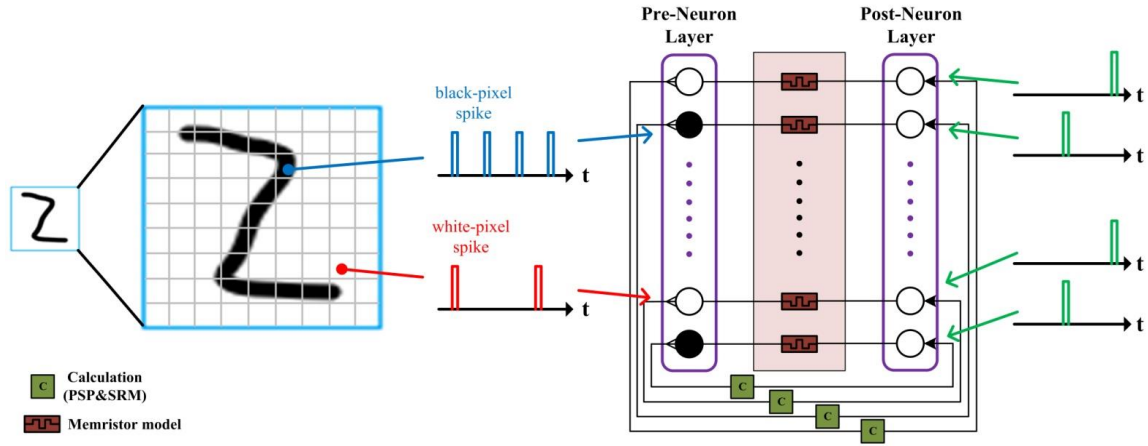

Fig. 4 The entire structure of STDP based memristive feedforward neural network

During test process, one random image is encoded into voltage signals and inputted to the well-trained networks, so the output currents are different due to the different values of memristors. If the sum of output currents is large enough, in other words, the input test image has a high matching rate with the well-trained networks, the input test image corresponds to the right category.

*B. non-STDP based Feedforward Structure*

From formula (3), we have:

$$M(t) = k_1 - k_2 \cdot Area_{spike} \tag{6}$$

$$dW = dG = d(\frac{1}{M(t)}) = d(\frac{1}{k_1 - k_2 \cdot Area_{spike}}) \tag{7}$$

where $Area_{spike} = \int_{-\infty}^{t} I(t)dt$.

From above, it can be seen that the updating weights have a linear relationship with the area of spikes. Hence, if the encoding mode of black and white spikes depends on different spikes heights and firing frequencies, we can get non-STDP based unsupervised memristive feedforward neural networks.

Inspired by it, we construct a non-STDP based memristive feedforward neural network as Fig. 5. First, the black and white pixels are encoded into high and low spikes with the different firing frequencies, respectively. The firing frequencies of black spikes are higher than white spikes'. Different from STDP based structures, the height of output spikes is between black and white spikes, which can be obtained by a time controller rather a calculation module combing with PSP and SRM. The time controller, which can make the output ports change whenever the state of input changes, is designed in this system. Through the time controller, the frequencies output spikes corresponding to black spikes are the same as input black spikes and the output spikes corresponding to white spikes are in the same way.

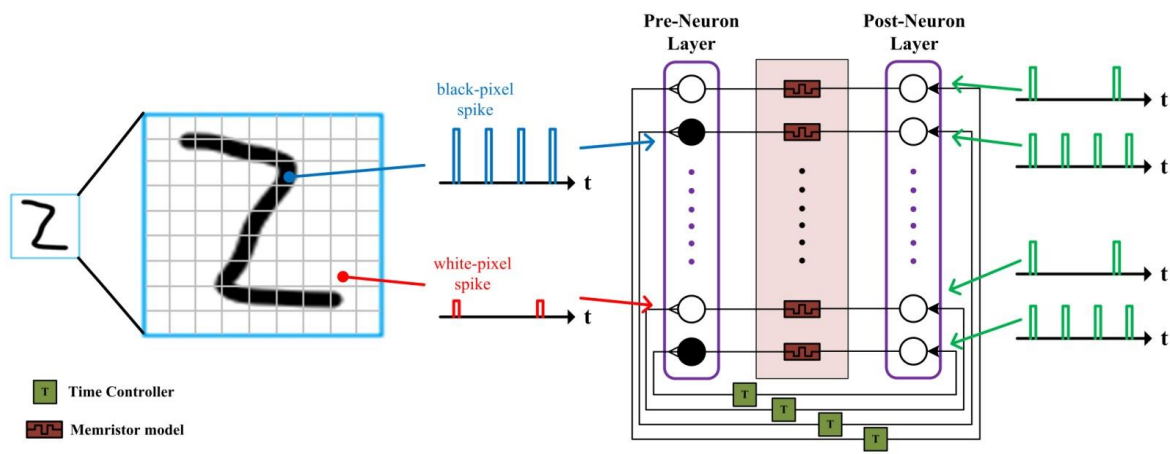

Fig. 5 The entire structure of non-STDP based memristive feedforward neural network

During the training process, the memristors corresponding to black spikes will be applied with spikes from left to right and the spikes through memristors corresponding to white spikes will be right to left. Hence, the black and white pixels will have different synaptic weights after training process. By this way, memristive characteristics can also be involved in our non-STDP method. Likewise, a fixed threshold for memristors is set to end the training process. The test process is the same as STDP based algorithm.

Compared with STDP based feedforward structure, the non-STDP feedforward structure only changes the encoding model and the way to get the output spikes. These two structures have the same numbers of layers, neurons, various modules, and memristors, which lays a good foundation for the comparison between STDP based and non-STDP based MNNs.

*C. Experiment Comparison and Analysis*

To carry out comparison of STDP based and non-STDP based memristive feedforward neural networks, a series of datasets from 3*3 pixels with 3categories (3*3*3) to 9*9 pixels with 5 categories (9*9*5) [18] are chosen as Fig. 6. Due to its advantages in parallelism and ease of operation, Intel FPGA Stratix V: 5SGXEA7N2F45C2 is chosen as the hardware platform to test networks' performance. To show the advantages of non-STDP based memristive feedforward neural network, some hardware optimizations including simplification, linearization, and pipeline design are applied in STDP based memristive feedforward neural network to achieve its better performance on hardware platform.

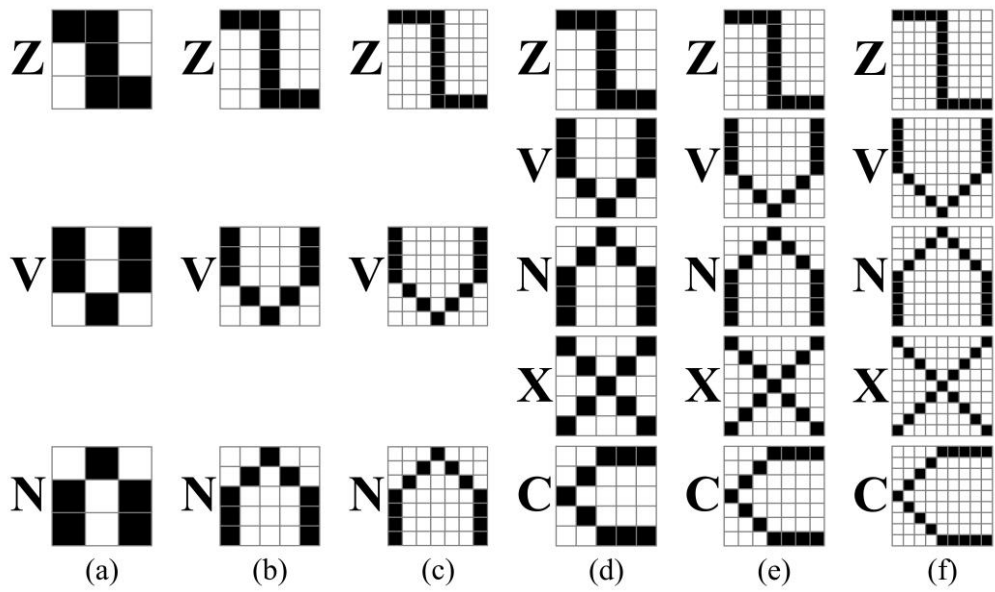

Fig. 6 A series of training and test datasets whose network scale is (a) 3*3*3, (b) 5*5*3, (c) 7*7*3, (d) 5*5*5, (e) 7*7*5, and (f) 9*9*5

The comparison consists of pattern recognition accuracy, convergence speed of training, hardware resource occupancy, and processing speed with the expansion of network scale, which is shown as Table I and Fig. 7.

TABLE I

COMPARISON OF MEMRISTIVE FEEDFORWARD STRUCTURE

| Network scale | | 3*3*3 | 5*5*3 | 7*7*3 | 5*5*5 | 7*7*5 | 9*9*5 |
|---|---|---|---|---|---|---|---|
| Pattern recognition accuracy | STDP | 100% | 100% | 99.9% | 99.8% | 99.5% | 99.1% |
| | non-STDP | 100% | 100% | 99.9% | 99.7% | 99.5% | 99.2% |
| Convergence speed (Number of training cycles) | STDP | 2 | 2 | 2 | 3 | 3 | 3 |
| | non-STDP | 2 | 2 | 2 | 3 | 3 | 3 |
| Hardware resource occupancy (ALMs) | STDP | 199 | 395 | 516 | 869 | 1309 | 1917 |
| | non-STDP | 185 | 337 | 390 | 750 | 1039 | 1826 |
| Processing speed (MHz) | STDP | 270.12 | 250.69 | 248.51 | 256.54 | 243.07 | 247.52 |
| | non-STDP | 274.06 | 260.82 | 259.04 | 261.76 | 255.32 | 251.41 |

It can be seen that STDP based and non-STDP based MNNs have similar pattern recognition accuracy and convergence speed for each network scale. Hence, non-STDP based MNNs have the same advantages as STDP based networks since memristive characteristics are also involved in our non-STDP algorithm.

In terms of hardware resource occupancy and process speed, non-STDP based feedforward algorithm has fewer hardware resources and higher processing speed. With the expansion of network scale from 3*3*3 to 9*9*5, non-STDP based algorithm keeps its edge because its mechanism is much simpler without following STDP on hardware. From above, it can be seen non-STDP MNNs have better hardware performance, which means better engineering compatibility.

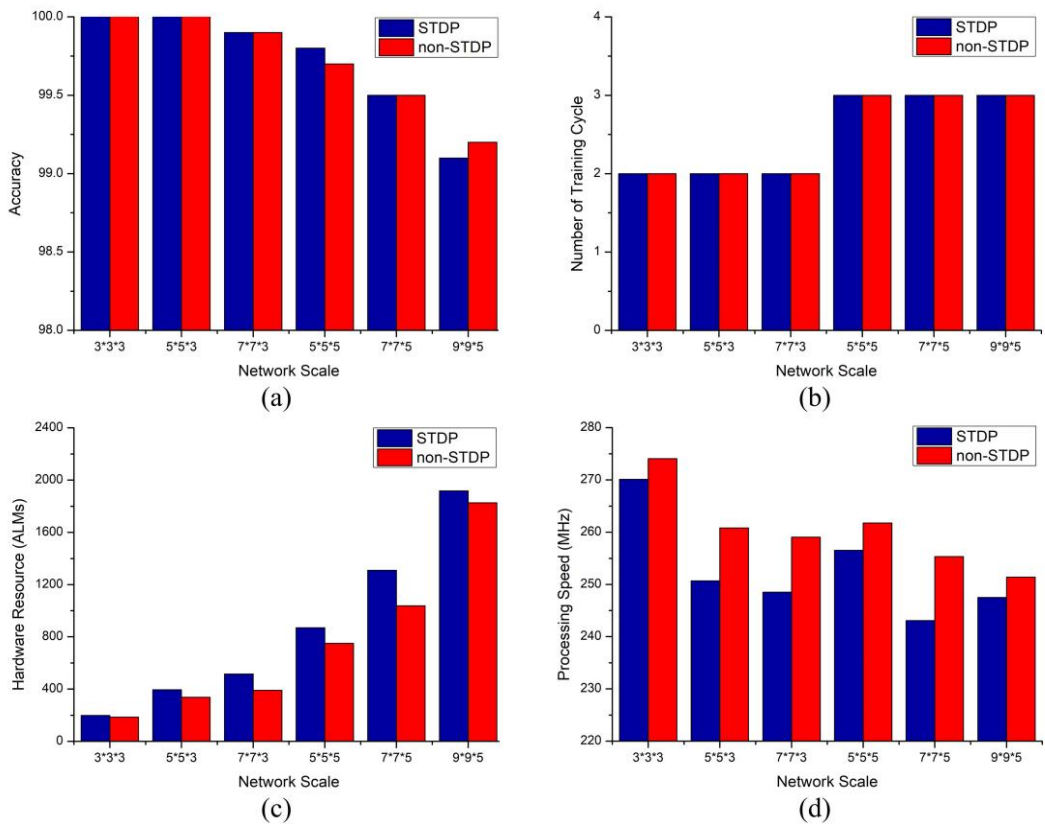

Fig. 7 Memristive feedforward structure's comparison including (a) pattern recognition accuracy, (b) convergence speed, (c) hardware resource occupancy, and (d) processing speed

III. MEMRISTIVE CROSSBAR STRUCTURE

In this section, a STDP based and a non-STDP based unsupervised memristive crossbar structures are built for comparison and analysis.

*A. STDP based Crossbar Structure*

Similar to feedforward structure, STDP based crossbar structure needs STDP, PSP, and SRM to form a learning system. The entire structure of STDP based memristive crossbar is shown as Fig. 8. First, the original images are encoded into spikes and the output spikes are got through calculation module including PSP and SRM. Then the crossbar structure will process training and weight updating following STDP unsupervised learning rule. We set a fixed threshold for each memristor to end training process. During the training process, the synaptic weights of memristors corresponding to black pixels in the diagonal from the top left to the bottom right will be the largest. Hence, after one random image is inputted to the well-trained crossbar during test process, the output current from these memristors will be larger than others if the input test image corresponds to the right category.

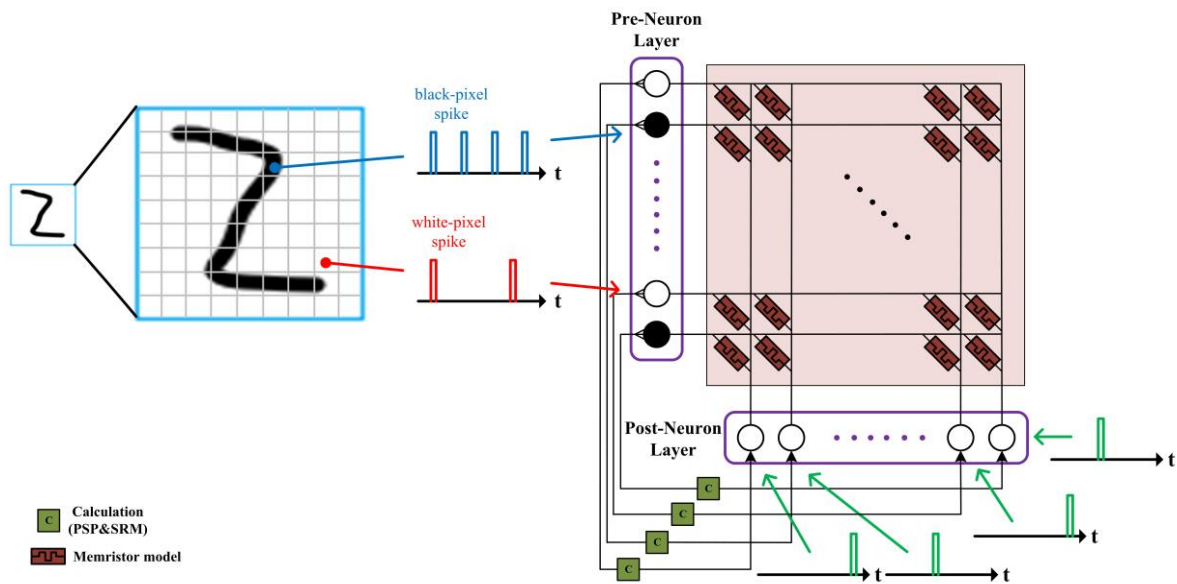

Fig. 8 The entire structure of STDP based memristive crossbar

*B. non-STDP based Crossbar Structure*

The entire structure of non-STDP based memristive crossbar is shown as Fig. 9. The non-STDP based crossbar also uses time controller to replace calculation module in STDP based crossbar. Similar to non-STDP feedforward neural networks, the black and white pixels are encoded into high and low spikes with the different firing frequencies, respectively, and the height of output spikes is between black and white spikes. During the training process, the synaptic weights of memristors corresponding to black pixels in the diagonal will be larger than others because the memristors corresponding to black pixels will be applied with spikes from left to right. Likewise, a fixed threshold for memristors to end the training process is set. The test process is the same as STDP based memristive crossbar structure.

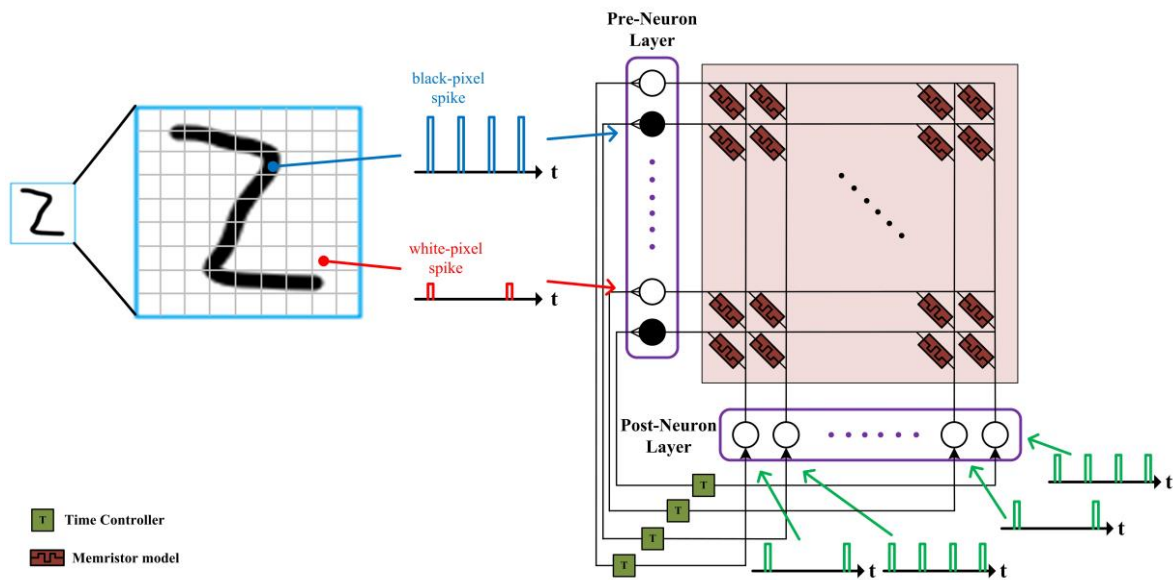

Fig. 9 The entire structure of non-STDP based memristive crossbar

Compared with STDP based crossbar, the non-STDP crossbar structure uses different encoding model and way to get the output spikes. These two structures have the same numbers of neurons, memristors, and various modules, so the performance of the two structures can be compared fairly.

*C. Experiment Comparison and Analysis*

A series of dataset from 3*3*3 to 9*9*5 and Intel FPGA Stratix V: 5SGXEA7N2F45C2 are chosen as dataset and hardware platform to test the performance of STDP based and non-STDP based memristive crossbar, respectively. Some hardware optimizations, such as simplification, linearization, and pipeline design, are applied in STDP based memristive crossbar to achieve its better performance, which can better show the advantages of non-STDP based memristive crossbar on hardware.

Comparison includes pattern recognition accuracy, convergence speed of training, hardware resource occupancy, and processing speed with the expansion of network scale, which is shown as Table II and Fig. 10.

TABLE II

COMPARISON OF MEMRISTIVE CORSSBAR STRUCTURE

| Network scale | | 3*3*3 | 5*5*3 | 7*7*3 | 5*5*5 | 7*7*5 | 9*9*5 |
|---|---|---|---|---|---|---|---|
| Pattern recognition accuracy | STDP | 100% | 100% | 100% | 99.9% | 99.6% | 99.3% |
| | non-STDP | 100% | 100% | 100% | 99.9% | 99.5% | 99.3% |
| Convergence speed (Number of training cycles) | STDP | 2 | 2 | 2 | 3 | 3 | 3 |
| | non-STDP | 2 | 2 | 2 | 3 | 3 | 3 |
| Hardware resource occupancy (ALMs) | STDP | 751 | 4706 | 17438 | 4789 | 17626 | 47056 |
| | non-STDP | 667 | 4475 | 16907 | 4512 | 17185 | 46027 |
| Processing speed (MHz) | STDP | 224.67 | 218.02 | 214.51 | 218.36 | 209.72 | 203.82 |
| | non-STDP | 227.22 | 223.69 | 216.07 | 220.06 | 213.42 | 208.16 |

From above, it can be seen that STDP based and non-STDP based crossbar have similar performance in pattern recognition accuracy and convergence speed with the expansion of network scale from 3*3*3 to 9*9*5. That means the non-STDP crossbar also has the MNN's advantages including high accuracy and convergence speed in pattern recognition. Further, because the mechanism of non-STDP based crossbar is much simpler, the hardware resource occupancy of non-STDP based crossbar is fewer and the processing speed is higher, which shows that non-STDP crossbar has the better hardware performance and is more suitable for practical engineering applications. The experimental results are consistent with the results of STDP based and non-STDP based memristive feedforward neural networks.

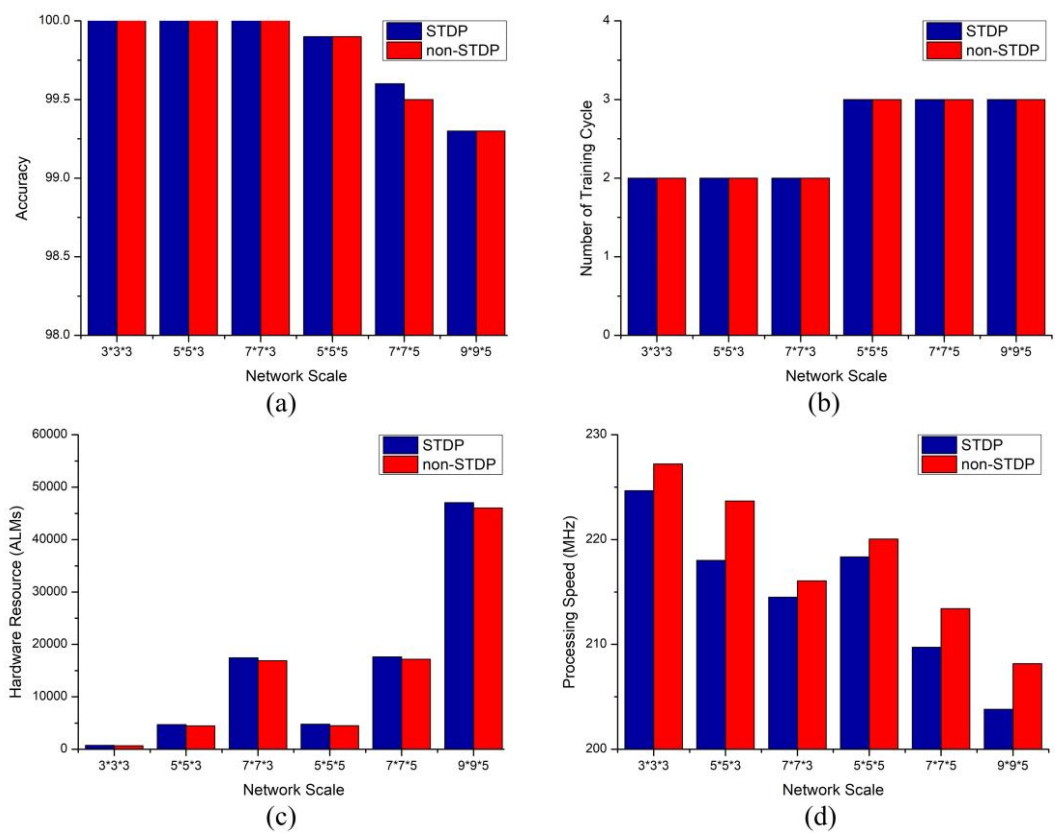

Fig. 10 Memristive crossbar structure's comparison including (a) pattern recognition accuracy, (b) convergence speed, (c) hardware resource occupancy, and (d) processing speed

## IV. CONCLUSION

Through the comparison between STDP based and non-STDP based MNNs with feedforward and crossbar structures, a phenomenon is pointed out that non-STDP based MNNs have advantages in hardware performance including hardware resource occupancy and processing speed remaining the similar pattern recognition accuracy and convergence speed with STDP based MNNs. Although STDP based memristive neural networks are in line with biological learning mechanism, the non-STDP based memristive structures show stronger hardware performance and engineering compatibility, due to its simple mechanism. That provides a new way for memristive neural networks' engineering applications.